\documentclass[aps,prapplied,reprint,groupedaddress,showpacs,showkeys]{revtex4-1}
\usepackage{amsfonts}
\usepackage{amssymb}
\usepackage{amsmath}
\usepackage{graphicx}
\usepackage{dcolumn}
\usepackage{bm}
\usepackage{units}
\usepackage{multirow}
\usepackage{CJKutf8}
\usepackage{subfigure}
\usepackage{color}
\usepackage{url}
\usepackage[colorlinks,linkcolor=red,anchorcolor=green,citecolor=blue]{hyperref}
\providecommand{\U}[1]{\protect\rule{.1in}{.1in}}
\usepackage{array}
\newcommand{\PreserveBackslash}[1]{\let\temp=\\#1\let\\=\temp}
\newcolumntype{C}[1]{>{\PreserveBackslash\centering}p{#1}}
\newcolumntype{R}[1]{>{\PreserveBackslash\raggedleft}p{#1}}
\newcolumntype{L}[1]{>{\PreserveBackslash\raggedright}p{#1}}
\begin{document}

\title{Intrinsic Picosecond Magnetic Switching Mechanism Assisted by an Electric Field in a Synthetic Antiferromagnetic Structure}

\author{Lei Wang (\begin{CJK}{UTF8}{gbsn}王蕾\end{CJK})}
\email{wanglei.icer@xjtu.edu.cn}
\affiliation{Center for Spintronics and Quantum Systems, State Key Laboratory for Mechanical Behavior of Materials, Xi'an Jiaotong University, No.28 Xianning West Road, Xi'an, Shaanxi, 710049, China}
\author{Runzi Hao}
\affiliation{Center for Spintronics and Quantum Systems, State Key Laboratory for Mechanical Behavior of Materials, Xi'an Jiaotong University, No.28 Xianning West Road, Xi'an, Shaanxi, 710049, China}
\author{Tai Min}
\email{tai.min@mail.xjtu.edu.cn}
\affiliation{Center for Spintronics and Quantum Systems, State Key Laboratory for Mechanical Behavior of Materials, Xi'an Jiaotong University, No.28 Xianning West Road, Xi'an, Shaanxi, 710049, China}
\date{\today}

\begin{abstract}
The processional switching mechanism governs magnetic switching in magnetic tunnel junctions (MTJs) in the sub-nanosecond range, which limits the application of spin transfer torque magnetic random access memory (STT-MRAM) in the ultrafast region. In this paper, we propose a new picosecond magnetic switching mechanism in a synthetic antiferromagnetic (SAF) structure using the adjustable Ruderman-Kittel-Kasuya-Yosida (RKKY) interaction controlled by an external electric field (E-field). It is shown that along with the sign change of the RKKY interaction in the SAF structure with an external E-field, the critical switching current density can be significantly reduced by one order of magnitude compared to that of a normal MTJ design at 100 ps; thus, this novel STT-MRAM can be written with a very low switching current density to avoid the MTJ breakdown problem and reduce the writing energy. To understand the physical origin of this abnormal phenomenon, a toy model is proposed in which the external-E-field-controlled sign change of the RKKY interaction in the SAF structure provides an extra contribution to the total energy that helps the
spins overcome the energy barrier and break the processional switching mechanism.
\end{abstract}

\maketitle

\section{Introduction}
Because of its nonvolatility and high density, spin-transfer-torque magnetic random access memory (STT-MRAM) has received extensive attention in both research \cite{apl1.3694270,2013JPhD.46g4001K,Kawahara2012} and industry \cite{6131602,6479128,4154269}. The core functional area in STT-MRAM is the magnetic tunnel junctions (MTJs) based on two magnetic layers sandwiching a tunnel barrier, in which a high tunnel magnetoresistance ratio (TMR) \cite{PhysRevB.63.054416,PhysRevB.63.220403,2004NatMa.3.862P} is used to read the data bits and the spin transfer torque
\cite{1996JMMM.159L.1S,PhysRevB.54.9353} from a polarized charge current is used to
write. However, because the angular momentum carried by electrons is generally limited, a large charge current is always required to switch the magnetization \cite{PhysRevLett.80.4281,Myers867}, which produces considerable unexpected Joule heating and results in an instability in the MTJs. More disadvantageously, when the writing process becomes faster ($<$ 10 ns), the critical charge current for switching increases exponentially based on processional switching mechanics \cite{PhysRevB.62.570,Diao2007}, which can even break down the tunnel barrier in the MTJs in the sub-1-ns region \cite{7573362}. Thus, until now, STT-MRAM still has not been competitive with L1/2 - static random-access memory (SRAM).

\begin{figure}[tp]
	\includegraphics[width=0.9\columnwidth]{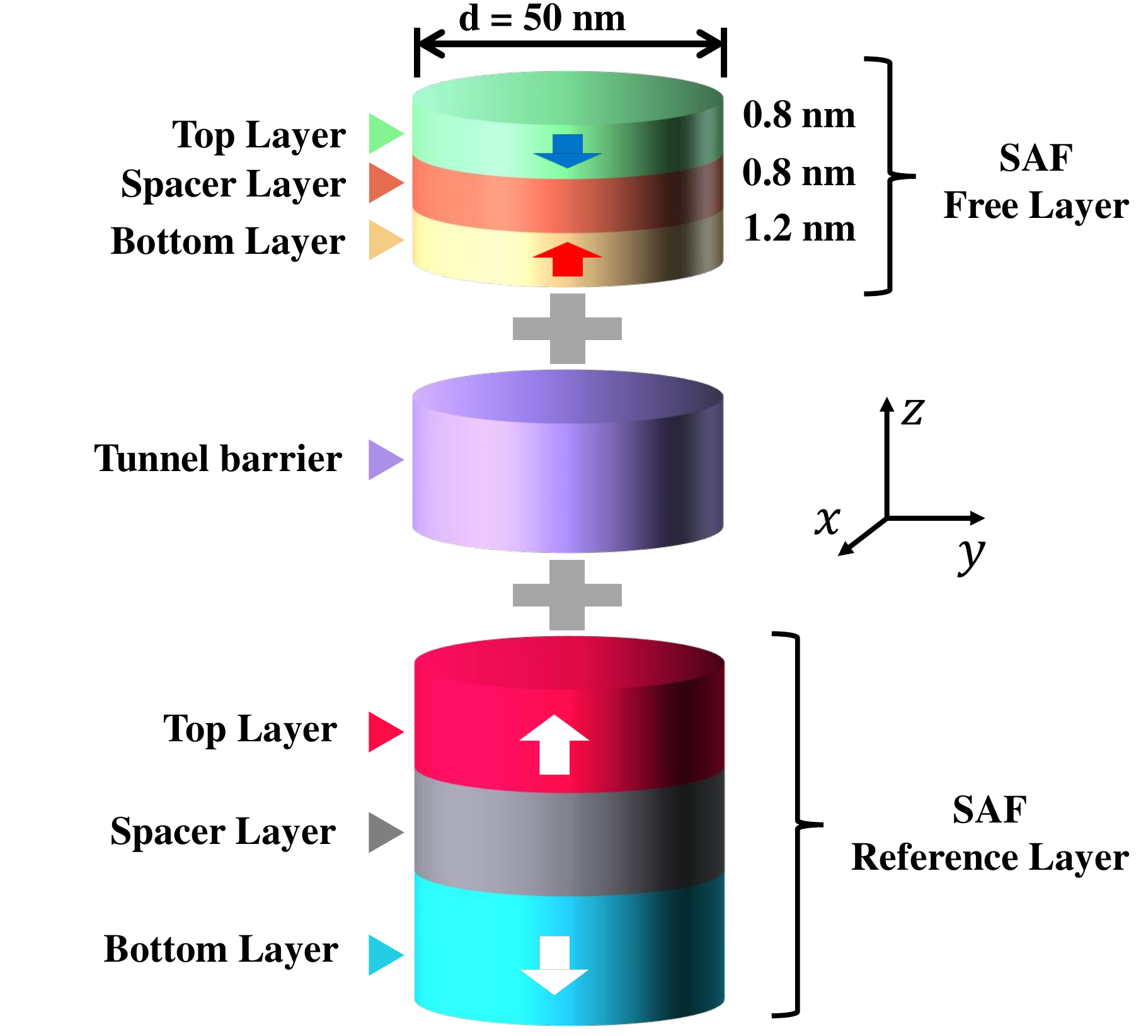}
\caption{Sketch model of the newly designed spin transfer torque magnetic random access memory (STT-MRAM), in which we use multiple synthetic antiferromagnetic (SAF) layers as a magnetic free layer in a traditional MTJ instead of general ferromagnetic metals. Additionally, the size of the SAF free layer used in the simulations is displayed in the figure.
The TMR is determined mainly by the tunneling effect through the tunnel barrier; therefore, the magnetic order between the bottom layer of the SAF free layer and the top layer of the SAF reference layer is the key element of the memory.}
	\label{fig1}
\end{figure}

Many new designs have been proposed in recent years to solve the above limitations of STT-MRAM in terms of the high energy consumption and breakdown problems in MTJs
induced by the large critical switching current density for a sub-10-ns switching time: e.g., a unique magnetic switching path design to reduce the critical switching current \cite{PhysRevB.97.144416} and spin orbit torque in a heavy metal and ferromagnetic (FM) metal interface to enhance the charge-spin-transfer efficiency \cite{2012Sci.336.555L,PhysRevB.94.174434}, which can also be applied to the interfaces between topological insulators (TIs) and magnetic thin films \cite{PhysRevLett.119.077702,8442225,8614499} for improved performance. Moreover, the observed giant interface spin Hall effect \cite{Wang.2016}, enhanced strong magnetic anisotropy of a heavy metal inset layer \cite{8408732,2018NatCo.9.671W}, atomic modification of the free layer interfaces \cite{PhysRevApplied.9.011002}, interface-based Rashba effect on the spin motive force \cite{PhysRevLett.108.217202} and spin mixing conductance \cite{PhysRevB.99.045421}, thermal-assisted spin transfer torque \cite{PhysRevLett.107.176603,2015PNAS.112.6585P,PhysRevMaterials.2.104403}, voltage-controlled magnetic anisotropy (VCMA) \cite{7835620,7172542} and adjustable perpendicular magnetic anisotropy (PMA) \cite{Adma.zhao} all contribute to increasing the efficiency of magnetic switching.

Recently, it has been proven that the ground state of a synthetic antiferromagnet can be changed from an antiferromagnetic (AFM) state to an FM state by an external electric field (E-field) \cite{2018NatCo.Yang} via tuning the sign of the Ruderman–Kittel–Kasuya–Yosida (RKKY) interaction \cite{PhysRev.96.99,kasuya,PhysRev.106.893,PhysRevB.44.7131}. This novel effect has inspired us to design a particularly special MTJ by replacing the traditional FM free layer with an SAF free layer and introducing an E-field-controlled AFM-to-FM phase transition to assist magnetic switching in the sub-1-ns region.

In this paper, we use micromagnetic simulations to study the magnetic switching of an SAF free layer with the help of an external E-field. The model and method used are given in Sec.~\ref{mm}, and the critical switching current density is presented in Sec.~\ref{cri} for both the the FM and SAF free layers for a comparison, where the critical switching current can be reduced by one order of magnitude in the 100 ps region. A toy model is proposed to explain these significant results. In Sec.~\ref{ap}, the critical switching current under an asynchronous pulse is studied for possible use in applications, and a summarized conclusion is given in Sec.~\ref{con}.

\section{Model and Method}
\label{mm}
A sketch model of the MTJ in our proposed SAF-based STT-MRAM is shown in Fig.~\ref{fig1}, which consists of an SAF free layer and an SAF reference layer sandwiching a tunnel barrier (e.g., MgO or $\mathrm{Al_2O_3}$). In this case, the storage bits are determined by the order (parallel or antiparallel) between the magnetization of the bottom layer of the SAF free layer and the top layer of the SAF reference layer, because the TMR in the whole device is dominated by the magnetic layer closed to the tunnel barrier. Because the direction of the magnetization in the SAF reference layer is fixed in both the read and write processes, the information of the SAF free layer is much more important in our study. Thus, we only use the SAF free layer to perform our simulations by injecting a polarized charge current to reproduce the function of the SAF reference layer and the tunnel barrier.

To investigate the spin dynamic process of the SAF free layer, simulations are carried out using the Object Oriented Micromagnetic Framework (OOMMF) \cite{ommff} code. In detail, we set up a nanopillar with an in-plane diameter of $d=50\ nm$, and the thickness of the SAF layer, $t=2.2~nm$, consists of two magnetic CoFeB layers ($t_{bottom}=1.2~nm$ and $t_{top}=0.8~nm$ for the bottom and top layer, respectively) and one nonmagnetic Ru ($t_{Ru}=0.8~nm$) layer. To balance the speed and accuracy of the simulations, the SAF free layer is discretized into a lattice of rectangular cells, and the size of every single cell is $2~nm\times2~nm\times0.4~nm$.

The dynamics of the spins are governed by the Landau-Lifshitz-Gilbert-Slonczewski (LLGS) \cite{PhysRevLett.102.067206,gilbert,landau,PhysRevB.70.172405} equation, which is:
\begin{eqnarray}
\frac{d\mathbf{m}}{d\tau}=-\gamma\mathbf{m}\times\mathbf{H}_{eff}+\alpha\mathbf{m}\times\frac{d\mathbf{m}}{d\tau}+\Gamma_{STT}
\end{eqnarray}
where $\mathbf{m}$ is the direction of the magnetization, $\tau$ is time, $\gamma$ is the gyromagnetic ratio, $\mathbf{H}_{eff}$ is the effective magnetic field and the damping constant $\alpha=0.01$ is used for CoFeB \cite{2011JAP.Liu,2013ApPhL.Devolder}. In addition, the spin transfer torque $\Gamma_{STT}$ generally comes from the injected polarized charge current, written as:
\begin{eqnarray}
\Gamma_{STT}=\gamma\beta\epsilon(\mathbf{m}\times\mathbf{m}_p\times\mathbf{m})-\gamma\beta\epsilon'\mathbf{m}\times\mathbf{m}_p
\end{eqnarray}
where $\beta=\frac{\hbar \mathrm{J}}{\vert e\vert \mu_0 t M_s}$, $\hbar$ is the reduced Planck constant, $\mathrm{J}$ is the charge current density, $e$ is the electron charge, $\mu_0$ is the vacuum permeability, $t$ is the thickness of the SAF free layer and the saturation magnetization of the magnetic CoFeB layer in the SAF free layer is $M_s=1.26\times10^6\ A/m$ \cite{2010NatMa.Ikeda}; in addition,
\begin{eqnarray}
\epsilon=\frac{P\Lambda^2}{(\Lambda^2+1)+(\Lambda^2-1)(\mathbf{m}\cdot\mathbf{m}_p)}
\end{eqnarray}
where $P=0.93$ is the polarization of the charge current with polarized direction $\mathbf{m}_p$, and we use $\Lambda = 1$ to remove the dependence of $\epsilon$ on $\mathbf{m}\cdot\mathbf{m}_p$ to make the STT isotropic. As the ratio of the field-like STT to the Slonczewski STT $\epsilon'/\epsilon$ in MgO-based MTJs varies from 0.1 to 0.3 \cite{2010APSMARL37010O,2008NatPh.Sankey,2008NatPh.Kubota,2008NatPh.Deac}, we set the secondary spin transfer term $\epsilon'=0.07$ to have an ordinary ratio of $\epsilon'/\epsilon=0.15$.

In addition to the general parameters of the materials in the SAF shown above, the total energy of the SAF free layer includes several parts, e.g., the Heisenberg exchange energy $\mathrm{E_{ex}}$ with the Heisenberg exchange coefficients $A=30\ pJ/m$ from CoFeB \cite{PhysRevB.83.054420}, the demagnetizing energy $\mathrm{E_{de}}$, and the anisotropy energy $\mathrm{E_{an}}=KV$ with the effective magnetic anisotropy constants $K$ and the volume $V$ of the SAF. However, we know that the magnetic anisotropy in MgO-based MTJs mainly arises from the interface; thus, we only use typical interface magnetic anisotropy constants in our calculations, which are $K_{bott}^i=1.44~erg/cm^2$ and $K_{top}^i=0.96~erg/cm^2$ \cite{2010NatMa.Ikeda} for the bottom layer and top layer of the SAF free layer, respectively. Here, the different interface magnetic anisotropy constants are chosen to make the bottom layer of the SAF much more stable because it is more important for storage, as described previously. In this case, the thermal stability can be obtained by \cite{2010NatMa.Ikeda,APL1.4858465}
\begin{eqnarray}
\begin{split}
\Delta=&\frac{K_{bott}^i/t_{bott}-\mu_0M_s^2/2}{k_B\mathcal{T}}V_{bott} \\
       &~~~~~~~~+\frac{K_{top}^i/t_{top}-\mu_0M_s^2/2}{k_B\mathcal{T}}V_{top}
\end{split}
\end{eqnarray}
where $V_{bott}$ and $V_{top}$ are the volumes of the bottom layer and top layer of the SAF structure, respectively, $k_B$ is the Boltzmann constant, and $\mathcal{T}$ is the temperature. Therefore, we have $\Delta\simeq199$ when $\mathcal{T}=300~K$, which
indicates very good thermal stability to have a more than 10 years retention of
the data \cite{4160113}.

Another important energy term in our simulations is the RKKY interaction, which can be described by adding an extra energy term in the calculations:
\begin{eqnarray}
\mathrm{E_{RKKY}}=\int_{i\in V} E_i~dV
\end{eqnarray}
where
\begin{eqnarray}
E_i=\sum_{j\in V}\frac{\sigma(1-\mathbf{m}_i\cdot\mathbf{m}_j)}{\delta_{ij}}
\label{ei}
\end{eqnarray}
is the density of the exchange energy of cell $i$ in one magnetic layer in the SAF structure
relative to all matching cells $j$ in the other layer, $\sigma$ represents the RKKY coefficients between the two magnetic layers and $\delta_{ij}$ is the discretization cell size in the direction from cell $i$ towards cell $j$.

With this approximation, one may notice that the two magnetic layers in the SAF structure will be ferromagnetically coupled when $\sigma>0$ and antiferromagnetically coupled when $\sigma<0$. Thus, the E-field-controlled AFM-to-FM phase transition in the SAF structure \cite{2018NatCo.Yang} can be governed by changing the sign of the RKKY coefficients $\sigma$ in the micromagnetic simulations \cite{8482460}. In this sense, the applied E-field pulses are generated by changing $\sigma$ at defined times in our calculations. Because the critical switching current of the SAF structure is quite robust to different $\sigma$ values, as shown in previous results \cite{8482460}, we typically chose $\sigma=2~erg/cm^2$ and $\sigma=-2~erg/cm^2$ to represent the simulation with and without the E-field, respectively.
\section{Critical switching current}
\label{cri}
\begin{figure}[tp]
	\includegraphics[width=\columnwidth]{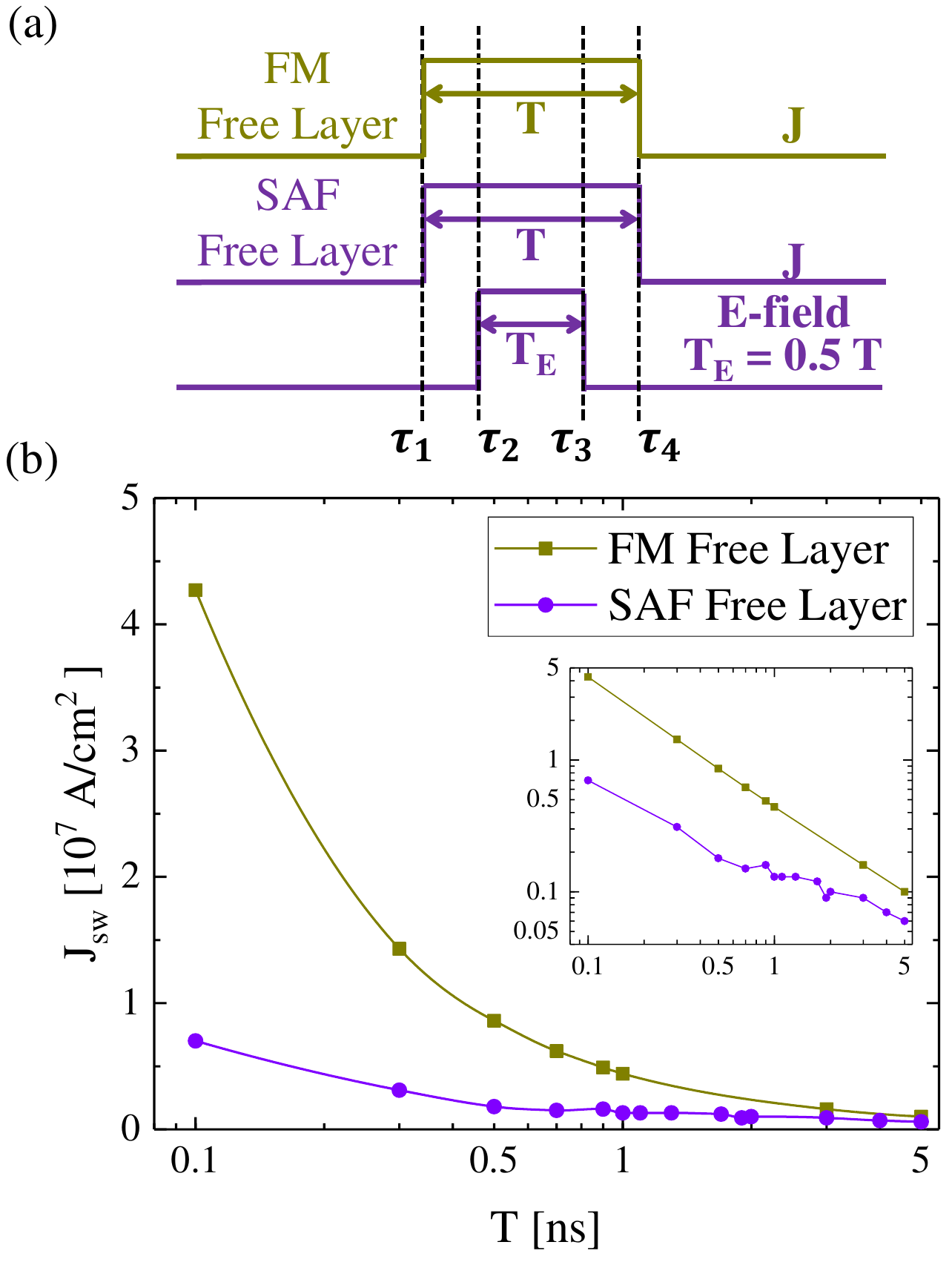}
\caption{(a) Sketch map of the applied charge current pulse $\mathrm{J}$ with pulse width $\mathrm{T}$ for both the FM metal free layer and SAF free layer and the corresponding external E-field pulse with $\mathrm{T_E=0.5T}$ for the SAF free layer only. Here, we also show the typical time points ($\tau_i,i\in\{1,2,3,4\}$) for further analysis. (b) The critical switching current density $\mathrm{J_{sw}}$ versus the current pulse width $\mathrm{T}$ for the normal FM free layer and the SAF free layer. The inset shows the same data on a log-log scale for a distinct comparison.}
	\label{fig2}
\end{figure}
It is well known that the speed of storage is limited by the writing process. Thus, we mainly investigate the magnetic switching of the SAF free layer, in which the charge current pulse width $\mathrm{T}$ and critical switching current $\mathrm{J_{sw}}$ are the most important parameters. Fig.~\ref{fig2} (a) shows a sketch of the charge current pulse and E-field pulse for the SAF free layer, and conventionally, we set the E-field pulse width to 
$\mathrm{T_E=0.5T}$ and put the E-field in the middle of the charge current pulse. Additionally, we mark four typical time points $\tau_{i,i\in\{1,2,3,4\}}$ where the charge current and E-field
pulse are turned on and off in Fig.~\ref{fig2} (a) for further detailed analysis. In contrast, the same charge current pulse is applied without an E-field pulse to a normal FM free layer, and a similar thermal stability factor $\Delta\simeq199$ of the FM free layer is chosen for a credible comparison.

\begin{figure*}[tp]
	\includegraphics[width=1.6\columnwidth]{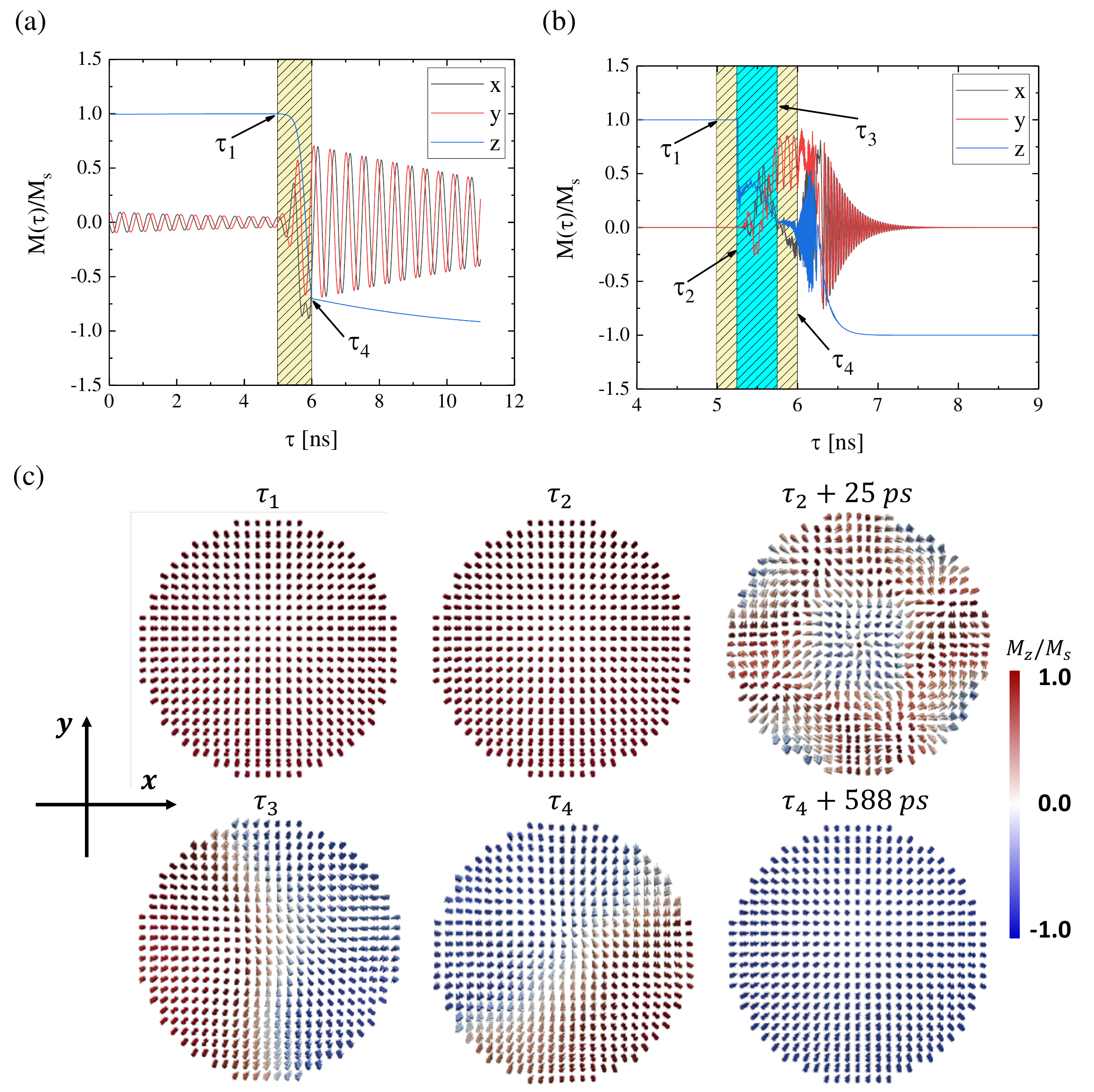}
\caption{The time-dependent normalized magnetization of the FM free layer (a) and the SAF free layer (b) with $\mathrm{T}=1$ ns from Fig.~\ref{fig2}. $\mathrm{M}(\tau)$ is obtained by averaging the whole FM layer and the bottom layer of the SAF differently, $x$, $y$, and $z$ represent the three components of the magnetization, and $\tau_{i},i\in\{1,2,3,4\}$ represent the typical time points, as displayed in Fig.~\ref{fig2} (a). (c) shows the visible spins in the bottom layer of the SAF free layer at the typical time points $\tau_i$, while the arrows point to the direction of each magnetization and the color
indicates the scalar magnitude of the out-of-plane components ($M_z$) according to color bar.}
	\label{fig3}
\end{figure*}

The calculated critical switching current density $\mathrm{J_{sw}}$ versus charge current pulse width $\mathrm{T}$ is plotted in Fig.~\ref{fig2} (b) for both the FM and SAF free layers. We also plot the same data on a log-log scale in the inset of Fig.~\ref{fig2} (b) for more details. The critical switching current density for the FM free layer increases exponentially, as expected from a previous publication \cite{Diao2007}, which is determined by processional switching mechanics \cite{PhysRevB.62.570}. However, when considering the SAF free layer, the critical switching current density decreases by almost one order of magnitude at $\mathrm{T}=0.1~ns$ when the effect of the sign change of the RKKY interaction is introduced by applying the E-field. Additionally, the critical switching current
density is still three times smaller at $\mathrm{T}=0.5~ns$, which makes the SAF free layer design much more energy-efficient in the sub-1-ns region as the Joule heating depends quadratically on the charge current. When increasing the charge current,  unexpected Joule heating may destroy the stability of the MTJ, and the high switching current can also break down the inside tunnel barrier, which is why the traditional STT-MRAM cannot operate in the sub-1-ns region \cite{7573362}. In this sense, from Fig.~\ref{fig2} (b), we can conclude that the speed of the SAF-based STT-MRAM still works at $\mathrm{T}=0.5~ns$ at least because the critical switching current density of the SAF free layer at $\mathrm{T}=0.5~ns$ is very close to that of the FM free layer at $\mathrm{T}=3~ns$, and researchers have already made traditional STT-MRAM applicable at $\mathrm{T}=3~ns$ \cite{7573412}. Thus, using the SAF-based STT-MRAM is at least six times faster than the traditional STT-MRAM, which makes it possible to replace the L1/2-SRAM at the sub-10-nm technology node \cite{8510672,6479128}.

\begin{figure*}[tp]
	\includegraphics[width=1.6\columnwidth]{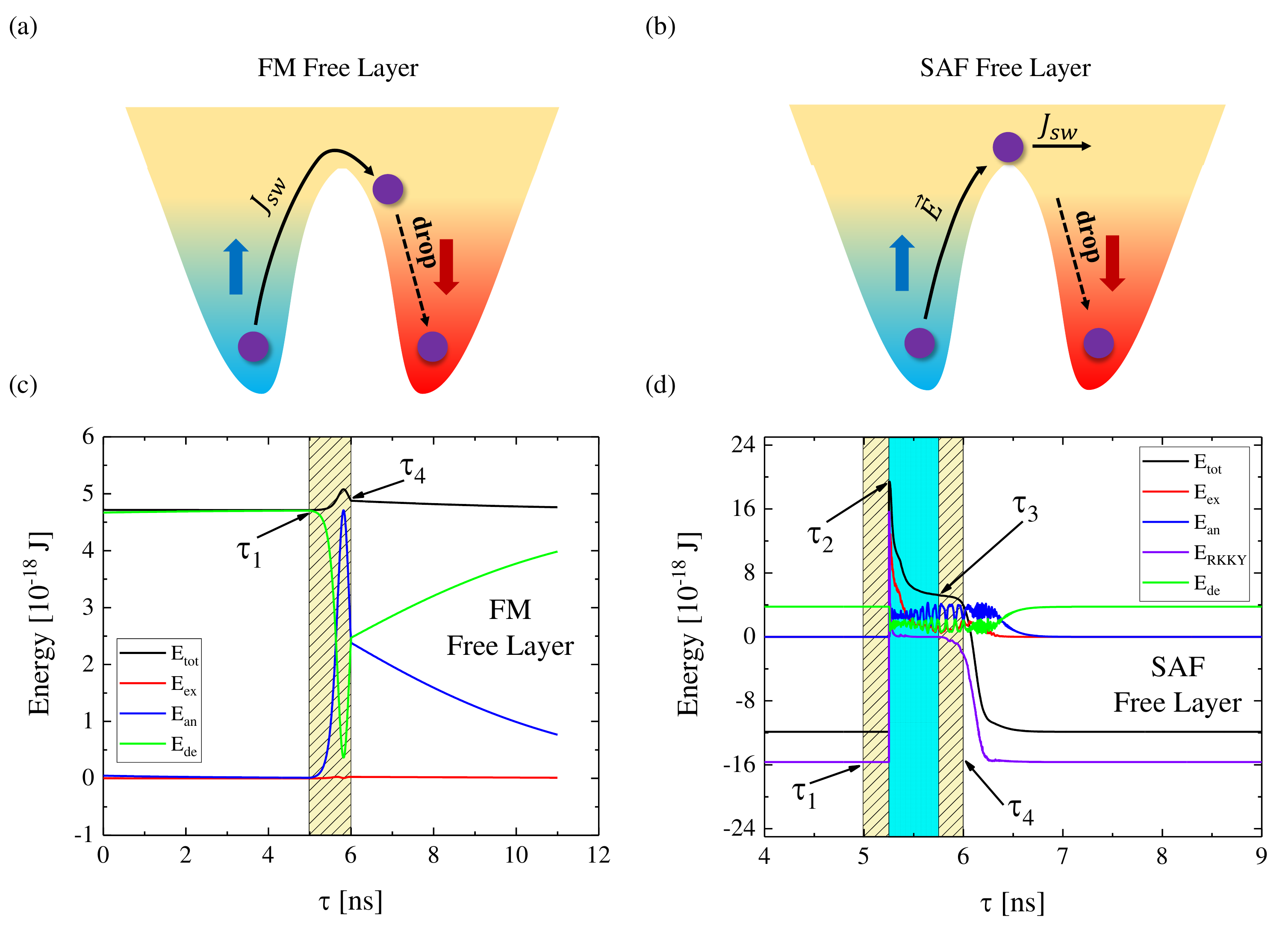}
\caption{The switching process of the FM free layer (a) and the SAF free layer (b) when $\mathrm{T}=1~ns$. In detail, the magnetization in the FM layer passes through the barrier between spin up and down only by the critical switching current density $\mathrm{J_{sw}}$, but for the SAF free layer, the external E-field provides an extra contribution to the
excitation of the spin from the ground state to an excitation state higher than the barrier, and the critical switching current density $\mathrm{J_{sw}}$ is now only a perturbation
 that is used to control the spin relaxation. (c) and (d) are the calculated energy versus time ($\tau$) for the FM free layer and the SAF free layer, where the total energy is $\mathrm{E_{tot}=E_{ex}+E_{an}+E_{de}+(E_{RKKY})}$, $\mathrm{E_{ex}}$ is the Heisenberg exchange energy, $\mathrm{E_{an}}$ is the anisotropy energy, $\mathrm{E_{de}}$ is the demagnetizing energy, and an extra RKKY interaction term $\mathrm{E_{RKKY}}$ is only considered in the SAF free layer.}
	\label{fig4}
\end{figure*}

In addition to the application of the SAF-based STT-MRAM design, the underlying physical origin of the E-field-assisted ultrafast switching in the SAF free layer still needs to be studied. Therefore, we focus on a typical charge current pulse width $\mathrm{T}=1~ns$ to investigate the dynamics of the FM and SAF free layers. As shown in Fig.~\ref{fig3} (a), we examine the ordinary processional switching mechanics in the FM free layer, in which the $z$ component (perpendicular to the interfaces in the MTJ) of the magnetization switches from 1.0 to -1.0 after applying the charge current at $\tau_1$, while the $x$ and $y$ components periodically rotate around the $z$-axis throughout the entire evolution time. However, in contrast, the magnetization in the SAF layer exhibits completely different behavior, as shown in Fig.~\ref{fig3} (b), where only the spins in the bottom layer of the
SAF structure have been considered for convenience. It can be seen that almost nothing happens when the charge current is applied at $\tau_1$; however, all three components ($x$, $y$, and $z$) of the magnetization exhibit a drastic change in a very short time when the E-field is turned on at $\tau_2$, erratically oscillate until a little time delay after turning off the charge current at $\tau_4$, and finally rotate normally, similar to the FM layer. This abnormal phenomenon of the dynamics of the magnetization provides strong evidence that the switching mechanics of the SAF free layer are far from the processional switching in the FM free layer, which should be the reason why the critical switching current in the SAF free layer can be reduced significantly, as shown in Fig.~\ref{fig2} (b).

\begin{figure*}[tp]
	\includegraphics[width=1.6\columnwidth]{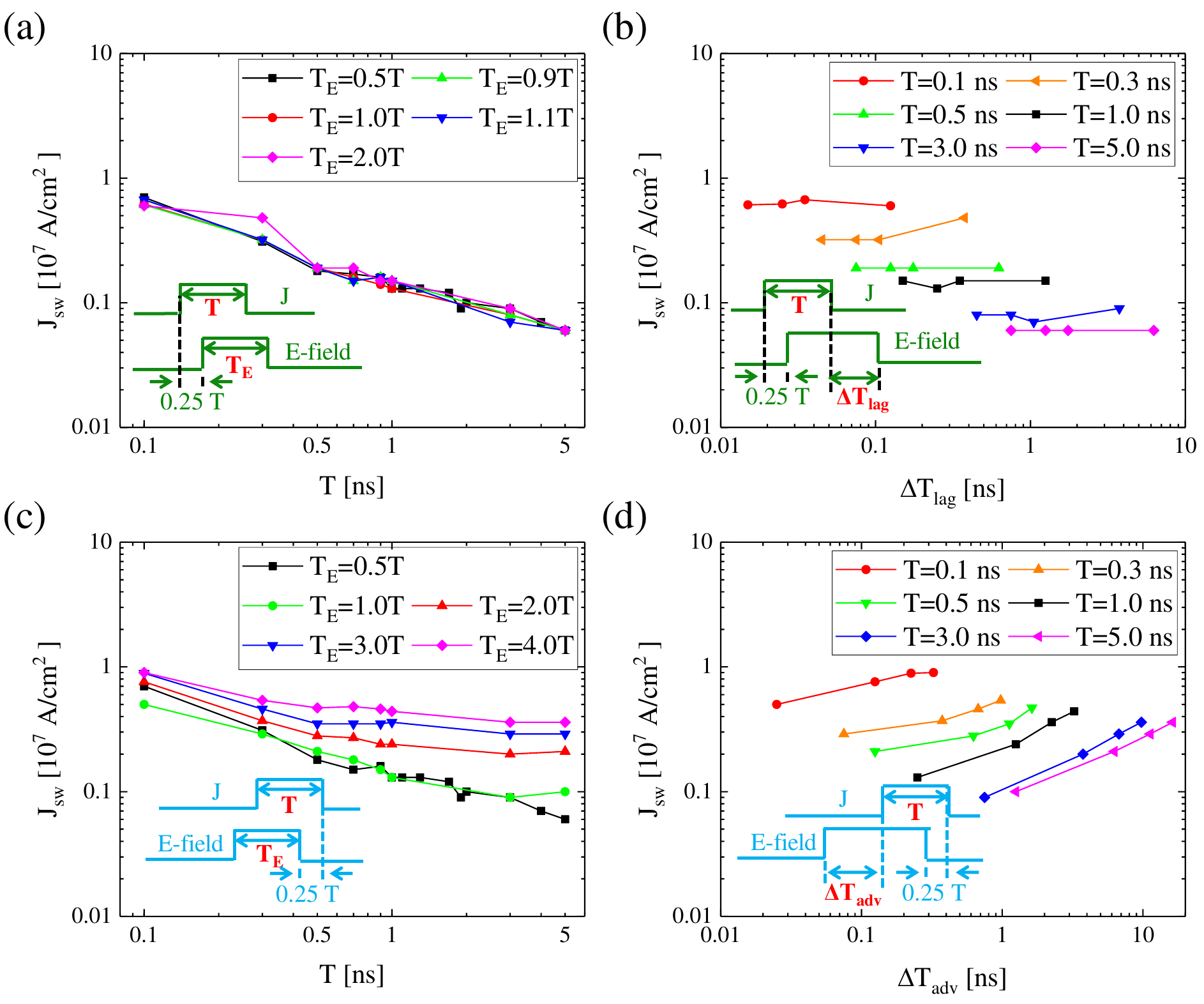}
\caption{The critical switching current density $\mathrm{J_{sw}}$ with an asynchronous charge current and E-field pulse, where the insets are the corresponding sketch maps of the detailed parameters of the current and E-field pulse; e.g., (a) and (c) plot $\mathrm{J_{sw}}$ as a function of the charge current pulse width $\mathrm{T}$ for various $\mathrm{T_E}$ of the E-field in both the lagging and advance cases, and (b) and (d) plot $\mathrm{J_{sw}}$ versus the lagging time and advance time between the E-field and the charge current for various $\mathrm{T}$.}
	\label{fig5}
\end{figure*}

For a more visible analysis, the texture of the spins in the bottom layer of the SAF free layer at some typical time points are shown in Fig.~\ref{fig3} (c), where the arrows point to the direction of every spin and the color stands for the $M_z$ component in particular. We can see that the charge current does not produce distinct spins until the E-field is turned on at $\tau_2$, and after 25 ps, the texture of the spins breaks down, which makes the uniform single domain change to a vertex-like multidomain. Then, the domains point in different
directions ($z$ or $-z$), start to compete with each other, and finally point to $-z$ (blue blue) with the help of the polarized charge current until the charge current is turned off at $\tau_4$. After approximately 588 ps, the whole layer switched from red ($\mathrm{m}\parallel z$) to blue ($\mathrm{m}\parallel-z$).

Based on all of these points, we propose a toy model as shown in Figs.~\ref{fig4} (a) and (b) for a physical understanding of this E-field-assisted ultrafast magnetic switching mechanics in the SAF free layer. Generally, there is an energy barrier between the two magnetic states, up and down, which determines how much energy is needed to switch the magnetization. As shown in Fig.~\ref{fig4} (a), for the normal FM free layer, only the applied charge current will supply enough angular momentum to help the magnetization overcome the barrier. However, for the SAF free layer (Fig.~\ref{fig4} (b)), when turning on the E-field, the total energy will increase very suddenly as the RKKY interaction coefficient $\sigma$ changes from $-2~erg/cm^2$ to $2~erg/cm^2$, according to Eq.~(\ref{ei}). This extra energy excites the magnetization from the ground state to an excitation state that is higher than the energy barrier; thus, the applied charge current only needs to provide a
perturbation to control the magnetization relaxation to an expected state (down in the figure) by its spin polarization and current direction. In this sense, the critical switching current density $\mathrm{J_{sw}}$ of the SAF free layer will not need to be as large as that in the FM free layer.

More convincing data support for the toy model are plotted in Figs.~\ref{fig4} (c) and (d), in which we show the total energy $\mathrm{E_{tot}}$ of the FM free layer and SAF free layer, respectively; moreover, the total energy is separated based on the different contribution terms, e.g., the Heisenberg exchange energy $\mathrm{E_{ex}}$, demagnetizing energy $\mathrm{E_{de}}$ and anisotropy energy $\mathrm{E_{an}}$ for both the FM and SAF free layers and an extra RKKY energy $\mathrm{E_{RKKY}}$ term only for the SAF free layer. From Fig.~\ref{fig4} (c), it can be seen that the Heisenberg exchange energy almost does not change during the whole evolution time, which indicates
uniform rotation of the magnetization in the FM free layer, with $\mathrm{E_{ex}}\propto\mathbf{m}_i\cdot\mathbf{m}_j$; in other words, the layers still obey processional
switching mechanics. Additionally, when focusing on the total energy $\mathrm{E_{tot}}$, one can easily find that the energy overcomes the energy barrier slowly after applying the charge current at $\tau_1$ and finally passes through the barrier slightly before $\tau_4$. However, for the SAF free layer in Fig.~\ref{fig4} (d), the Heisenberg exchange energy
exhibits a very sharp curve at $\tau_2$ when the E-field is turned on, which also indicates multidomain switching, as shown in Fig.~\ref{fig3} (c). Fig.~\ref{fig4} (d) clearly shows that the sharp enhancement in the total energy at $\tau_2$ is mainly due to the narrow peak of the RKKY interaction $\mathrm{E_{RKKY}}$ (purple line). All these data agree with our previous assumptions in the toy model.

\section{Asynchronous Pulse}
\label{ap}
\begin{figure}[tp]
	\includegraphics[width=\columnwidth]{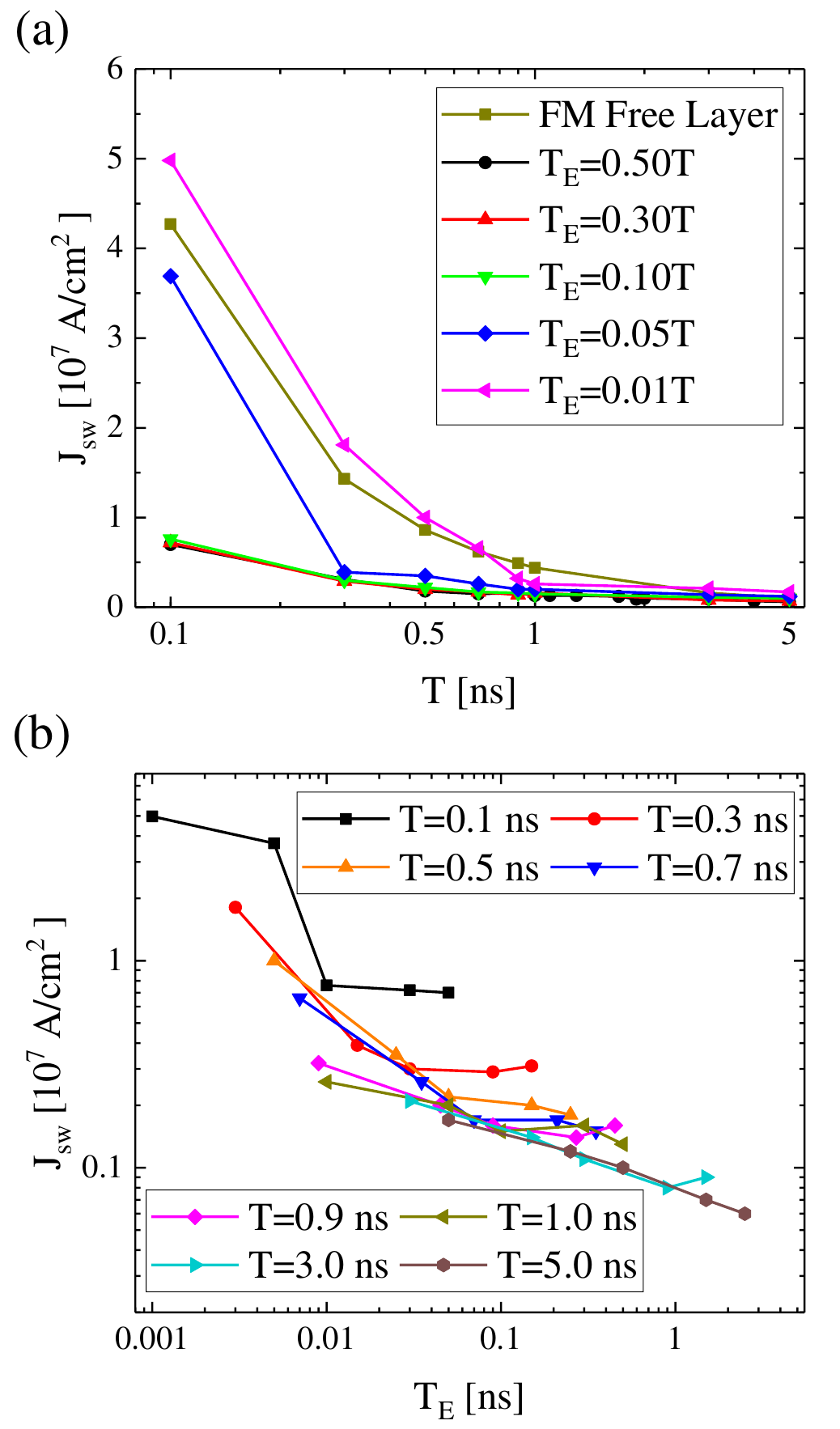}
\caption{(a) The critical switching current density $\mathrm{J_{sw}}$ versus the charge current pulse width $\mathrm{T}$ for various E-field pulse widths $\mathrm{T_E}$. (b)  $\mathrm{J_{sw}}$ versus the E-field pulse width $\mathrm{T_E}$ for various charge current pulse widths $\mathrm{T}$. Here, one should note that the current pulse and E-field pulse are applied similarly as shown in Fig.~\ref{fig2} (a); however, we keep $\tau_2-\tau_1=0.25\mathrm{T}$ for convenience in the simulations.}
	\label{fig6}
\end{figure}
Synchronization is always a difficult issue in circuit control. Therefore, the asynchronous condition between the charge current pulse and the E-field pulse should be a very important point in applications. Thus, we first increase the ratio between the E-field pulse width $\mathrm{T_E}$ and charge current pulse width $\mathrm{T}$ while fixing the turning-on times $\tau_1$ and $\tau_2$, as shown in the inset of Fig.~\ref{fig5} (a), to investigate the effect of the time delay of the E-field. The corresponding results of the critical switching current density $\mathrm{J_{sw}}$ versus $\mathrm{T}$ for various $\mathrm{T_E}$ are plotted in Fig.~\ref{fig5} (a), in which we find insensitive dependencies between $\mathrm{J_{sw}}$ and $\mathrm{T_E}$ for all the charge current pulse widths. These results can be understood within the framework of the previous toy model, where the function of the E-field is only used to excite the state of the SAF free layer and the charge current determines the magnetization relaxation; therefore, the time delay of the E-field will
not change the current physical picture. To show a more clear picture, we introduce the concept of the lagging time $\Delta \mathrm{T_{lag}}=\tau_3-\tau_4$ as shown in the insets of Fig.~\ref{fig5} (b) and plot $\mathrm{J_{sw}}$ as a function of $\Delta \mathrm{T_{lag}}$ for various $\mathrm{T}$. Even on a log-log scale, the critical switching current density is still robust with respect to the time delay of the E-field.

Similarly, as shown in the inset of Fig.~\ref{fig5} (c), we can increase the ratio between $\mathrm{T_E}$ and $\mathrm{T}$ while fixing the turning-off time of the E-field pulse $\tau_3$ and charge current pulse $\tau_4$ to investigate the effect of the time advance of the E-field. The calculated results are plotted in Fig.~\ref{fig5} (c), in which one can see that, unsurprisingly, while increasing $\mathrm{T_E}$, the critical switching current density $\mathrm{J_{sw}}$ increases with a positive correlation. We also introduce the concept of the advance time $\Delta \mathrm{T_{adv}}=\tau_1-\tau_2$ as shown in the insets of Fig.~\ref{fig5} (d) and show the relation between $\mathrm{J_{sw}}$ and $\Delta \mathrm{T_{adv}}$. Additionally, the unexpected increasing critical current appears in all
cases. This is because when the E-field turns on too far ahead of the charge current, the SAF free layer will have enough time to change from an AFM state to an FM state before the charge current provides any contribution; i.e., the applied charge will be used to pass through the energy barrier by itself without the help from the E-field, which is quite similar to that in the FM free layer. However, a smaller critical switching current is needed in applications with a high energy efficiency and fast read and write speed. Under these considerations, one should try to focus on turning on the E-field after the charge current ($\tau_1<\tau_2$) and relaxing the turning-off time $\tau_3$ of the E-field.

Moreover, according to our previous toy model, the most useful part of the E-field is the first few ps ($<$25 ps) for $\mathrm{T}=1~ns$, as displayed in Fig.~\ref{fig3} and
Fig.~\ref{fig4}. Therefore, it is unclear how short the E-field pulse widths $\mathrm{T_E}$ should be to realize our SAF free layer design for ultrafast reading and writing. Fig.~\ref{fig6} (a) plots the calculated critical switching current $\mathrm{J_{sw}}$ versus the charge current pulse width $\mathrm{T}$ for various normalized $\mathrm{T_E}$. It can be seen that, especially when $\mathrm{T}<1~ns$, $\mathrm{J_{sw}}$ increases very fast after $\mathrm{T_E}<0.1\mathrm{T}$ and becomes comparable to that in the FM free layer with $\mathrm{T_E}=0.01\mathrm{T}$. Conventionally, we can replot $\mathrm{J_{sw}}$ as a function of the absolute E-field pulse width $\mathrm{T_E}$ for various $\mathrm{T}$ in Fig.~\ref{fig6} (b), in which it can be found that $\mathrm{J_{sw}}$ increases significantly when decreasing $\mathrm{T_E}$ for all the charge current pulse widths $\mathrm{T}$. These results indicate that there exists a limitation for the E-field pulse width in the sense that the E-field needs sufficient time to break the single-domain texture in the SAF free layer. In addition, combined with the previous results under an asynchronous pulse, the best solution for this SAF free layer design involves using the lagging E-field, as shown in Fig.~\ref{fig5} (a).

\section{Conclusion}
\label{con}
In this paper, we perform micromagnetic simulations to study the magnetic switching dynamics in an SAF free layer under the assistance of an external E-field. The calculated results show that within the framework of the E-field-controlled AFM-to-FM phase transition in the SAF structure, the critical switching current density can be reduced by one order of magnitude at 100 ps; thus, our new design of the SAF-based STT-MRAM could potentially be used to replace L1/2-SRAM. To understand this significant phenomenon, the time-dependent dynamics have been investigated visibly, in which the single domain is broken, overcoming
the limitation of the processional switching mechanism. In addition, based on a detailed analysis of the energy evolution, we propose a toy model to describe the process of the E-field-assisted ultrafast switching mechanics; the model enables a concise physical picture where, in the SAF free layer, the E-field is used to excite the magnetic domain from the ground state to an energy level higher than the barrier, while the charge current
determines the magnetization relaxation.

For the possible technical use of the SAF-based STT-MRAM, we also investigate the effect of an asynchronous charge current and E-field pulse. We find that the critical switching current density is quite robust when the E-field pulse is longer than the charge current pulse; however, when the E-field pulse is applied ahead of the charge current pulse, the reduction in the critical switching current density becomes worse. Moreover, the E-field pulse cannot be reduced arbitrarily, because sufficient time is required to break the single domain in the SAF structure. Thus, in summary, the best strategy for applying the E-field is to turn on the E-field slightly slower than the charge current and to maintain the E-field as long as possible.

\begin{acknowledgments}
This work was supported by the National Key Research and Development Program of China (grant Nos. 2018YFB0407600, 2016YFA0300702 and 2017YFA0206202), the National Natural Science Foundation of China (grant No. 11804266) and Shaanxi Province Science and Technology Innovation Project (grant 2015ZS-02).
\end{acknowledgments}

\bibliographystyle{apsrev4-1}
\bibliography{SAF-STT}

\end{document}